\documentclass[aps,prc,preprint,tightenlines,groupedaddress,nofootinbib,
showpacs,preprintnumbers,amsmath,amssymb,superscriptaddress]{revtex4}
\usepackage{graphicx}% Include figure files
\usepackage{dcolumn}%Align table columns on decimal point
\usepackage{bm}% bold math
\usepackage{amsmath}
\begin{document}

\title{Pygmy dipole resonance as a constraint \\
       on the neutron skin of heavy nuclei}
\author{J. Piekarewicz}
\affiliation{Department of Physics, Florida State 
             University, Tallahassee, FL 32306}
\date{\today} 

\begin{abstract}
 The isotopic dependence of the isovector Pygmy dipole response in tin
 is studied within the framework of the relativistic random phase
 approximation. Regarded as an oscillation of the neutron skin against
 the isospin-symmetric core, the pygmy dipole resonance may place
 important constraints on the neutron skin of heavy nuclei and, as a
 result, on the equation of state of neutron-rich matter. The present
 study centers around two questions. First, is there a strong
 correlation between the development of a neutron skin and the
 emergence of low-energy isovector dipole strength? Second, could one
 use the recently measured Pygmy dipole resonance in ${}^{130}$Sn and
 ${}^{132}$Sn to discriminate among theoretical models? For the first
 question we found that while a strong correlation between the neutron
 skin and the Pygmy dipole resonance exists, a mild anti-correlation
 develops beyond ${}^{120}$Sn. The answer to the second question
 suggests that models with overly large neutron skins---and thus stiff
 symmetry energies---are in conflict with experiment.
\end{abstract}
\maketitle 

\section{Introduction}
\label{Sec:Intro}
The accurate determination of the neutron radius of a heavy nucleus
remains an unsolved nuclear-structure problem. The search for this
fundamental observable has been recently re-invigorated due to its
far-reaching influence on a host of interesting and seemingly
unrelated observables. For example, a correlation has been found
between the neutron radius of $^{208}$Pb and the binding energy of 
the valence neutrons in neutron-rich nuclei: models with small
neutron radii reach the neutron drip line sooner than those 
with larger radii~\cite{Todd:2003xs}. Further, an additional 
correlation was found between the neutron radius of ${}^{208}$Pb and 
the neutron radius of other heavy nuclei. This suggests that the 
measurement of the neutron radius of a single heavy nucleus 
could determine, or at least place strong constraints, on the neutron 
radius of a variety of nuclei. This could provide a boost to the atomic 
parity-violation program~\cite{Pollock:1992mv}. Many other physical 
observables also display a strong correlation with the neutron radius of 
${}^{208}$Pb. These include the equation of state of neutron-rich 
matter~\cite{Brown:2000}, isospin diffusion in heavy-ion 
collisions~\cite{Tsang:2004,Chen:2004si,Steiner:2005rd}, and the 
structure and dynamics of neutrons stars~\cite{Horowitz:2000xj,
Horowitz:2001ya,Horowitz:2002mb,Steiner:2004fi,Li:2005sr}

Attempts at mapping the neutron distribution have traditionally relied
on strongly-interacting probes. While highly mature and successful, it
is unlikely that the hadronic program will ever attain the precision
status that the electroweak program enjoys. This is due to the large
and controversial uncertainties in the reaction
mechanism~\cite{Ray:1985yg,Ray:1992fj}. The mismatch in our knowledge
of the proton radius in ${}^{208}$Pb relative to that of the neutron
radius provides a striking example of the current situation: while the
charge radius of ${}^{208}$Pb is known to better than
0.001~fm~\cite{Fricke:1995}, realistic estimates place the uncertainty
in the neutron radius at about 0.2 fm~\cite{Horowitz:1999fk}.

The enormously successful parity-violating program at the Jefferson
Laboratory~\cite{Aniol:2005zf,Aniol:2005zg} provides an attractive
electroweak alternative to the hadronic program. Indeed, the Parity
Radius Experiment (PREX) at the Jefferson Laboratory aims to measure
the neutron radius of $^{208}$Pb accurately (to within $0.05$~fm) and
model independently via parity-violating electron
scattering~\cite{Horowitz:1999fk}. Parity violation at low momentum
transfers is particularly sensitive to the neutron density because the
$Z^0$ boson couples primarily to neutrons. Moreover, the
parity-violating asymmetry, while small, can be interpreted with as
much confidence as conventional electromagnetic scattering
experiments. PREX will provide a unique observational constraint on
the thickness of the neutron skin of a heavy nucleus. We note that
since first proposed in 1999, many of the technical difficulties
intrinsic to such a challenging experiment have been met. For
example, during the recent activity at the Hall A Proton Parity
Experiment (HAPPEX), significant progress was made in controlling
helicity correlated errors~\cite{Michaels:2005}. Other technical
problems are currently being solved---such as the designed of a new
septum magnet---and a specific timeline has been provided to solve all
remaining problems within the next two years~\cite{Michaels:2005}.

While this important experiment gets off the ground, a significant
effort has been devoted to constrain the neutron radius of a heavy
nucleus by alternative means. One such effort uses nuclear giant
resonances to constrain bulk properties of infinite nuclear matter,
such as the compression modulus and the symmetry
energy~\cite{Piekarewicz:2002jd,Agrawal:2003xb,Vretenar:2003qm,
Piekarewicz:2003br,Colo:2004mj}. Knowledge of the density dependence
of the symmetry energy places important constraints on the neutron
skin of heavy nuclei. Indeed, the slope of the symmetry energy---a
quantity related to the pressure of neutron-rich matter---has been
shown to be strongly correlated to the neutron radius of heavy
nuclei~\cite{Brown:2000}. Yet the reliance on nuclear excitations
stems from the inability of existent ground-state observables to
constrain the density dependence of the symmetry energy. While
accurately calibrated models reproduce a variety of ground-state
observables, they still generate a large spread in the neutron skin 
of heavy nuclei~\cite{Brown:2000,Furnstahl:2001un}.

In this presentation we focus on the ``pygmy'' dipole resonance (PDR),
a soft nuclear mode that has the promise of constraining the neutron
skin of heavy nuclei. The timing of this work is motivated by a
pioneering experiment that has identified low-energy isovector dipole
strength in two unstable neutron-rich isotopes: ${}^{130}$Sn and
${}^{132}$Sn~\cite{Adrich:2005}.  Pictured as an oscillation of the
neutron skin of the neutron-rich nucleus against the isospin-symmetric
core, the emergence of low-energy strength is expected to follow
closely the development of the neutron skin.  The goals of the present
manuscript are therefore twofold: (a) to use the isotopic chain in tin
to search for a correlation between the development of a neutron skin
and the emergence of low-energy isovector dipole strength, and (b) to
use the recent experimental data~\cite{Adrich:2005} to discriminate
among theoretical models that, while accurately calibrated, yield
different predictions for the neutron skin of heavy nuclei.

Theoretical studies of the Pygmy dipole resonance are not new. The
possible existence of a new type of dipole oscillation in which the
neutron skin vibrates against the isospin-symmetric core dates back to
the early 1990s~\cite{Suzuki:1990,VanIsacker:1992}. Shortly after,
sophisticated mean-field (MF) plus random-phase-approximation (RPA)
approaches were developed and used to predict the distribution of
low-energy isovector dipole strength on a variety of neutron-rich
nuclei~\cite{Hamamoto:1996,Hamamoto:1998,Vretenar:2000yy,Vretenar:2001hs}.
More recently, some of these models have attained a level of
sophistication that surpass MF+RPA approaches by incorporation such
effects as pairing correlations and/or the coupling to more complex
configurations~\cite{Paar:2002gz,Tsoneva:2003gv,Sarchi:2004,Paar:2004gr}.

In the present contribution we compare for the first time relativistic
MF+RPA calculations against the experimental data of
Ref.~\cite{Adrich:2005}. The experimental data is used to discriminate
among accurately-calibrated models having a different density
dependence for the symmetry energy, such as
NL3~\cite{Lalazissis:1996rd,Lalazissis:1999} and
FSUGold~\cite{Todd-Rutel:2005fa}. The self-consistent MF+RPA approach
employed here neglects both pairing correlations and the coupling to
more complex configurations. Yet by direct comparison to some of the
most recent
studies~\cite{Paar:2002gz,Tsoneva:2003gv,Sarchi:2004,Paar:2004gr}---or
by the authors own admission---these effects do not affect the main
conclusions of the present work.

The manuscript has been organized as follows. In
Sec.~\ref{sec:formalism} we introduce some of the most basic details
of the MF+RPA formalism. For a more comprehensive discussion the
reader is referred to Ref.~\cite{Piekarewicz:2001nm}. In
Sec.~\ref{sec:results} results for the distribution of isovector
dipole strength in the neutron-even Sn-isotopes are presented. In the
same section we discuss the various effects that may be used to
elucidate the neutron skin of heavy nuclei and, correspondingly, the
equation of state of neutron-rich matter. We close by offering our
conclusions in Sec.~\ref{sec:conclusions}.

\section{Formalism}
\label{sec:formalism}

The starting point for the calculation of the nuclear response
is an interacting Lagrangian density of the following form:
\begin{widetext}
\begin{eqnarray}
&&
{\cal L}_{\rm int} =
\bar\psi \left[g_{\rm s}\phi   \!-\!
         \left(g_{\rm v}V_\mu  \!+\!
    \frac{g_{\rho}}{2}{\mbox{\boldmath $\tau$}}\cdot{\bf b}_{\mu}
                               \!+\!
    \frac{e}{2}(1\!+\!\tau_{3})A_{\mu}\right)\gamma^{\mu}
         \right]\psi \nonumber \\
                   && -
    \frac{\kappa}{3!} (g_{\rm s}\phi)^3 \!-\!
    \frac{\lambda}{4!}(g_{\rm s}\phi)^4 \!+\!
    \frac{\zeta}{4!}
    \Big(g_{\rm v}^2 V_{\mu}V^\mu\Big)^2 \!+\!
    \Lambda_{\rm v}
    \Big(g_{\rho}^{2}\,{\bf b}_{\mu}\cdot{\bf b}^{\mu}\Big)
    \Big(g_{\rm v}^2V_{\mu}V^\mu\Big) \;.
\label{Lagrangian}
\end{eqnarray}
\end{widetext}
The Lagrangian density includes an isodoublet nucleon field ($\psi$)
interacting via the exchange of two isoscalar mesons, a scalar
($\phi$) and a vector ($V^{\mu}$), one isovector meson ($b^{\mu}$),
and the photon ($A^{\mu}$)~\cite{Serot:1984ey,Serot:1997xg}. In
addition to meson-nucleon interactions the Lagrangian density is
supplemented by four nonlinear meson interactions with coupling
constants denoted by $\kappa$, $\lambda$, $\zeta$, and $\Lambda_{\rm
v}$.  The first three of these terms are responsible for a softening
of the equation of state of symmetric nuclear matter---at both normal
and high densities~\cite{Mueller:1996pm}. The last coupling constant
($\Lambda_{\rm v}$) induces isoscalar-isovector mixing and is
responsible for modifying the density dependence of the symmetry
energy~\cite{Horowitz:2000xj,Horowitz:2001ya}. As a result of the
strong correlation between the pressure of neutron-rich matter and the
neutron radius of heavy nuclei~\cite{Brown:2000,Furnstahl:2001un},
$\Lambda_{\rm v}$ may also be used to modify the neutron radius of
heavy nuclei.

\subsection{Ground-state Properties}
\label{groundstate}

The initial step in the study of nuclear excitations is the
calculation of ground-state properties. This is done by solving the
equations of motion associated with the above Lagrangian
self-consistently in a relativistic mean-field approximation. What
emerges from such a calculation is a set of binding energies, a
corresponding set of wave-functions, and a self-consistent mean-field
potential. From these wave-functions, ground-state densities and all
their corresponding moments ({\it e.g.,} root-mean-square radii) may
be extracted.

%%%%%%%%%%%%%%%%%%%%%%%%%%%%%%%%%%%%%%%%%%%%%%%%%%%%%%%%%%%%%%%%%%%%%%
\begin{figure}[ht]
\vspace{0.50in}
\includegraphics[width=5in,angle=0]{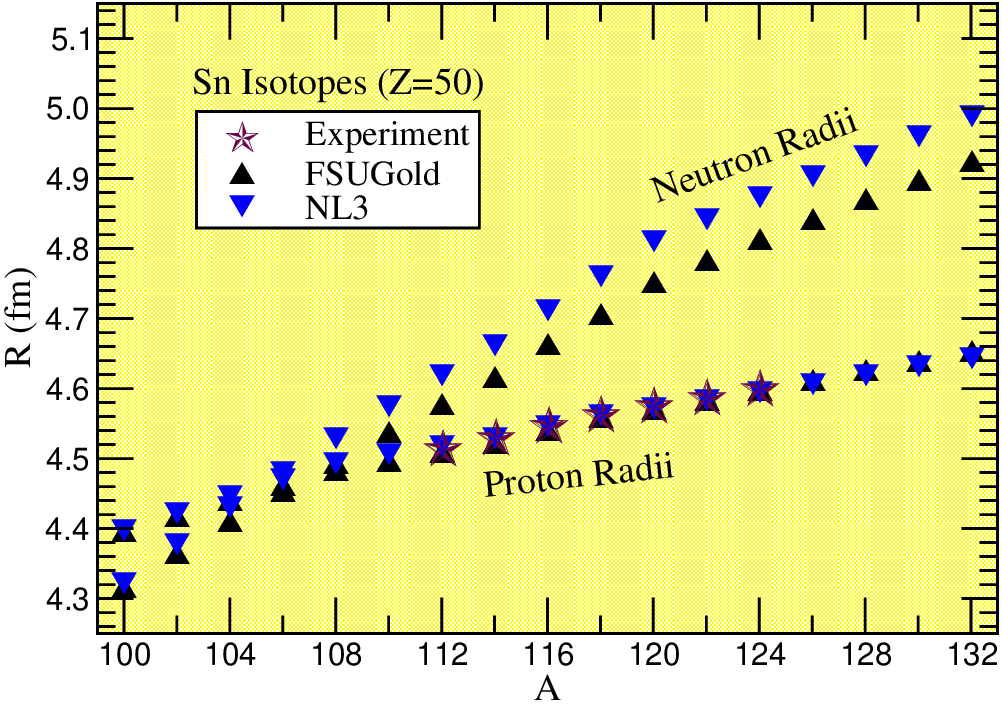}
\caption{Proton and neutron root-mean-square radii for a variety of
         even-neutron Sn-isotopes computed in a relativistic
         mean-field approximation using the FSUGold and NL3 parameter 
         sets. Experimental data, when available, is from 
         Ref.~\cite{DeJager:1987qc}.
}
\label{Fig1}
\end{figure}
%%%%%%%%%%%%%%%%%%%%%%%%%%%%%%%%%%%%%%%%%%%%%%%%%%%%%%%%%%%%%%%%%%%%%%

A typical example of such a mean-field procedure is provided in
Fig.~\ref{Fig1}, where point proton and point neutron root-mean-square
radii are displayed for all neutron-even isotopes of tin ranging from
the doubly-magic nucleus ${}^{100}$Sn to the doubly-magic nucleus
${}^{132}$Sn. When available, these calculations are compared against
experimental data. The same information alongside ground-state
energies is displayed in Table~\ref{Table1}.  Two
accurately-calibrated parameter
sets---NL3~\cite{Lalazissis:1996rd,Lalazissis:1999} and
FSUGold~\cite{Todd-Rutel:2005fa}---were employed in the calculation.
Masses and coupling constants for these two models have been listed in
Table~\ref{Table2}. Note that for the NL3 parameter set, both $\zeta$
and $\Lambda_{\rm v}$ have been set equal to zero. The absence of a
quartic vector coupling term ($\zeta$) should not come as a
surprise, as the variety of ground-state observables employed in the
calibration procedure (such as binding energies and charge radii) are
insensitive to the high-density component of the equation of
state. Yet the absence of the isoscalar-isovector coupling
$\Lambda_{\rm v}$---or any other term that may change the density
dependence of the symmetry energy---is significant. It suggests that
despite the vast amount of accurate ground-state data collected over
the years, it is not possible at present to pin down the neutron
radius of even a single heavy nucleus. Both Fig.~\ref{Fig1} and
Table~\ref{Table1} indicate that, when a comparison is possible, the
agreement between theory and experiment is excellent. Yet these two
models, although accurately calibrated, predict vastly different
neutron radii for the neutron-rich isotopes.

%%%%%%%%%%%%%%%%%%%%%%%%%%%%%%%%%%%%%%%%%%%%%%%%%%%%%%%%%%%%%%%%%
\begin{table}
\begin{tabular}{|c|c|c|c|c|c|c|c|c|}
 \hline
   & \multicolumn{3}{|c|}{FSUGold} &
     \multicolumn{3}{ c|}{NL3}     &
     \multicolumn{2}{ c|}{Experiment} \\
\hline
   A & $B/A$~(MeV) & $R_{p}$~(fm) & $R_{n}\!-\!R_{p}$~(fm)
     & $B/A$~(MeV) & $R_{p}$~(fm) & $R_{n}\!-\!R_{p}$~(fm)
     & $B/A$~(MeV) & $R_{p}$~(fm) \\
\hline
 100 & $8.244$    & $4.397$ & $-0.080$ 
     & $8.284$    & $4.397$ & $-0.076$ 
     & $8.249(4)$ &   ---   \\
 102 & $8.293$    & $4.419$ & $-0.053$
     & $8.312$    & $4.421$ & $-0.044$ 
     & $8.323(4)$ &   ---   \\
 104 & $8.345$    & $4.440$ & $-0.030$
     & $8.355$    & $4.445$ & $-0.015$ 
     & $8.383(2)$ &   ---   \\
 106 & $8.399$    & $4.462$ & $-0.009$
     & $8.396$    & $4.469$ & $\phantom{-}0.011$ 
     & $8.433(1)$ &   ---   \\
 108 & $8.456$    & $4.484$ & $\phantom{-}0.010$
     & $8.440$    & $4.493$ & $\phantom{-}0.035$ 
     & $8.468\phantom{(0)}$ &   ---   \\
 110 & $8.470$    & $4.497$ & $\phantom{-}0.041$
     & $8.455$    & $4.575$ & $\phantom{-}0.070$ 
     & $8.496\phantom{(0)}$ &   ---   \\
 112 & $8.487$    & $4.510$ & $\phantom{-}0.069$
     & $8.474$    & $4.517$ & $\phantom{-}0.102$ 
     & $8.514\phantom{(0)}$ & $4.516$   \\
 114 & $8.506$    & $4.523$ & $\phantom{-}0.094$
     & $8.495$    & $4.528$ & $\phantom{-}0.133$ 
     & $8.523\phantom{(0)}$ & $4.532$   \\
 116 & $8.496$    & $4.541$ & $\phantom{-}0.122$
     & $8.486$    & $4.545$ & $\phantom{-}0.167$ 
     & $8.523\phantom{(0)}$ & $4.549$   \\
 118 & $8.490$    & $4.559$ & $\phantom{-}0.148$
     & $8.480$    & $4.561$ & $\phantom{-}0.199$ 
     & $8.517\phantom{(0)}$ & $4.564$   \\
 120 & $8.477$    & $4.570$ & $\phantom{-}0.182$
     & $8.468$    & $4.572$ & $\phantom{-}0.238$ 
     & $8.505\phantom{(0)}$ & $4.577$   \\
 122 & $8.447$    & $4.584$ & $\phantom{-}0.200$
     & $8.446$    & $4.583$ & $\phantom{-}0.259$ 
     & $8.488\phantom{(0)}$ & $4.589$   \\
 124 & $8.420$    & $4.598$ & $\phantom{-}0.216$
     & $8.425$    & $4.595$ & $\phantom{-}0.278$ 
     & $8.467\phantom{(0)}$ & $4.601$   \\
 126 & $8.396$    & $4.612$ & $\phantom{-}0.231$
     & $8.407$    & $4.607$ & $\phantom{-}0.296$ 
     & $8.444\phantom{(0)}$ &   ---   \\
 128 & $8.375$    & $4.626$ & $\phantom{-}0.245$
     & $8.391$    & $4.619$ & $\phantom{-}0.314$ 
     & $8.417\phantom{(0)}$ &   ---   \\
 130 & $8.356$    & $4.640$ & $\phantom{-}0.259$
     & $8.377$    & $4.631$ & $\phantom{-}0.330$ 
     & $8.388\phantom{(0)}$ &   ---   \\
 132 & $8.334$    & $4.654$ & $\phantom{-}0.271$
     & $8.365$    & $4.644$ & $\phantom{-}0.346$ 
     & $8.355\phantom{(0)}$ &   ---   \\
\hline
\end{tabular}
\caption{Binding energy per nucleon, proton root-mean-square radii,
         and neutron-skin thickness for a variety of even-neutron 
         Sn-isotopes computed in a relativistic mean-field approximation 
         using the FSUGold and NL3 parameter sets. A center-of-mass 
	 correction of the form $(3/4)41A^{-4/3}$ has been applied to 
         the binding energy per nucleon. Experimental
         data is from Refs.~\cite{Audi:1995,DeJager:1987qc}.}
\label{Table1}
\end{table}
%%%%%%%%%%%%%%%%%%%%%%%%%%%%%%%%%%%%%%%%%%%%%%%%%%%%%%%%%%%%%%%%%

%%%%%%%%%%%%%%%%%%%%%%%%%%%%%%%%%%%%%%%%%%%%%%%%%%%%%%%%%%%%%%%%%
\begin{table}
\begin{tabular}{|l||c|c|c|c|c|c|c|c|}
 \hline
 Model & $m_{\rm s}$  & $g_{\rm s}^2$ & $g_{\rm v}^2$ & $g_{\rho}^2$
       & $\kappa$ & $\lambda$ & $\zeta$ & $\Lambda_{\rm v}$\\
 \hline
 \hline
 NL3     & 508.1940 & 104.3871  & 165.5854 &  79.6000
         & 3.8599 & $-$0.0159 & 0.0000 & 0.0000   \\
 \hline
 FSUGold & 491.5000 & 112.1996  & 204.5469 & 138.4701
         & 1.4203 & $+$0.0238 & 0.0600 & 0.0300   \\
\hline
\end{tabular}
\caption{Model parameters used in the calculations. The parameter
$\kappa$ and the inverse scalar range $m_{\rm s}$ are given in MeV.
The nucleon, omega, and rho masses are kept fixed at $M\!=\!939$~MeV,
$m_{\omega}\!=\!782.5$~MeV, and $m_{\rho}\!=\!763$~MeV, respectively.}
\label{Table2}
\end{table}
%%%%%%%%%%%%%%%%%%%%%%%%%%%%%%%%%%%%%%%%%%%%%%%%%%%%%%%%%%%%%%%%%

\subsection{Isovector Dipole Response}
\label{IsovDipResp}

The calculations presented here follow closely the formalism
developed in much greater detail in Ref.~\cite{Piekarewicz:2001nm}. It
is however advantageous to provide a brief summary of the main points
of the approach.  The isovector dipole response will be extracted from
the timelike component of the polarization tensor.  Consistent with
the assumption of a mean-field ground state, the polarization tensor
is computed using a relativistic MF+RPA formalism. Moreover, the
residual particle-hole interaction is consistent with the mean-field
potential employed to generate the ground state.  This, by itself and
with nothing else, guarantees both the decoupling of the spurious
strength from the RPA response as well as the conservation of the
vector current~\cite{Dawson:1990wp,Piekarewicz:2001nm}.

To start, we introduce the most general polarization tensor which is 
defined in terms of a time-ordered product of two arbitrary nucleon 
currents:
\begin{equation}
  i\Pi^{\alpha\beta}(x,y) =  \langle \Psi_{0}| 
  T \Big( \hat{J}^{\alpha}(x)\hat{J}^{\beta}(y)\Big) 
  |\Psi_{0}\rangle \;.
 \label{PiAlphaBeta}
\end{equation}
Here $\Psi_{0}$ denotes the exact nuclear ground state and
$\hat{J}^{\alpha}(x)$ is a one-body current operator of the 
form
\begin{equation}
  \hat{J}^{\alpha}(x) = \bar{\psi}(x)\Gamma^{\alpha}\psi(x) \;,
 \label{jalpha}
\end{equation}
where $\Gamma^{\alpha}$ is a matrix having an arbitrary Dirac and 
isospin structure. In the particular case of nuclear excitations 
of isovector dipole character, a simple operator of the following
form will be used to drive the transition:
%%%
\begin{equation}
 \Gamma^{\alpha}\rightarrow \Gamma^{0}_{3} \equiv \gamma^{0}\tau_{3} \;,
 \label{G03}
\end{equation}
where $\tau_{3}$ is a $2\times 2$ isospin matrix and for the
$\gamma$-matrices we have adopted the convention of
Ref.~\cite{Bjorken:1964}. Finally, due to the invariance under time
translation, it is convenient to rewrite the polarization tensor in
terms of the excitation energy $\omega$ as follows:
\begin{equation}
  \Pi^{\alpha\beta}(x,y) = \int_{-\infty}^{\infty} 
  \frac{d\omega}{2\pi}e^{-i\omega(x^{0}-y^{0})}
  \Pi^{\alpha\beta}({\bf x},{\bf y};\omega) \;.
 \label{PixyOmega}
\end{equation}

The isovector dipole response, denoted henceforth as $S_{L}(q,\omega)$, 
may now be extracted from the imaginary part of a suitable polarization 
tensor. That is, 
\begin{equation}
   S_{L}(q,\omega)=-\frac{1}{\pi}
   \Im\,\Pi^{00}_{33}({\bf q},{\bf q};\omega) \;,
 \label{S03}
\end{equation}
where the above labels refer to the isovector-timelike operator of
Eq.~(\ref{G03}) and $\Pi^{00}_{33}({\bf q},{\bf q}';\omega)$ denotes
the Fourier transform of $\Pi^{00}_{33}({\bf x},{\bf y};\omega)$.
Such an operator is capable of exciting all natural-parity states. 
To isolate the isovector dipole response one must project out
the $(J^{\pi}\!=\!1^{-};T\!=\!1)$ component of the polarization
tensor. Thus, a transition operator of the following form is used:
\begin{equation}
 \widehat{D}^{0}_{1\mu;3}(q,{\bf r})=
  j_{1}(qr)Y_{1\mu}(\hat{\bf{r}})\gamma^{0}\tau_{3}
  \mathop{\longrightarrow}_{(qr\ll 1)}
  \frac{1}{3}qrY_{1\mu}(\hat{\bf{r}})\gamma^{0}\tau_{3}\;.
 \label{IVDipole}
\end{equation}

In the mean-field approximation the nuclear polarization tensor may 
be written exclusively in terms of the nucleon mean-field propagator 
$G_{\rm MF}(x,y)$ as follows:
\begin{equation}
  i\Pi^{\alpha\beta}_{\rm MF}(x,y) = {\rm Tr}
  \Big(\Gamma^{\alpha}G_{\rm MF}(x,y)\Gamma^{\beta}G_{\rm MF}(y,x)\Big) \;,
 \label{PiAlphaBetaMF}
\end{equation}
where the nucleon propagator $G^{\alpha\beta}_{\rm MF}$ is defined, in 
analogy to Eq.~(\ref{PiAlphaBeta}), as a time-ordered product of two 
nucleon fields
\begin{equation}
  iG^{\alpha\beta}_{\rm MF}(x,y) =  \langle \Phi_{0}| 
  T \Big( \psi^{\alpha}(x)\bar{\psi}^{\beta}(y)\Big) 
  |\Phi_{0}\rangle \;.
 \label{GAlphaBeta}
\end{equation}
What defines the mean-field propagator is the replacement of the exact
ground state of the system $\Psi_{0}$ by its mean-field approximation
$\Phi_{0}$. In the mean-field approximation the spectral content of
the polarization tensor is both simple and
illuminating~\cite{Fetter:1971}.  The polarization tensor is an
analytic function of the excitation energy $\omega$---except for
simple poles located at the one-particle--one-hole excitations of the
mean-field system, with the residues at the individual poles yielding
the transition form-factors.

To build collectivity into the nuclear response, all single-particle 
excitations of the same spin and parity must be mixed via a residual 
particle-hole interaction. The collective response of the system to 
an external perturbation is then obtained as a solution of the 
following set of RPA (Dyson's) equations~\cite{Fetter:1971}:
\begin{equation}
  \Pi^{\alpha\beta}_{\rm RPA}({\bf q},{\bf q}';\omega) =
  \Pi^{\alpha\beta}_{\rm MF}({\bf q},{\bf q}';\omega) \!+\!
  \int\!\frac{d^3k}{(2\pi)^{3}}
       \frac{d^3k'}{(2\pi)^{3}}
  \Pi^{\alpha\lambda}_{\rm MF}({\bf q},{\bf k};\omega)           
   V_{\lambda\sigma}({\bf k},{\bf k}';\omega)
  \Pi^{\sigma\beta}_{\rm RPA}({\bf k}',{\bf q}';\omega) \;,
 \label{PiabRPA} 
\end{equation}
where $V_{\lambda\sigma}({\bf k},{\bf k}';\omega)$ is the
(momentum-space) residual particle-hole interaction. It should be
stressed that in order to preserve important symmetries of the
problem, the residual interaction must be consistent with the
interaction used to generate the mean-field ground
state~\cite{Dawson:1990wp,Piekarewicz:2001nm}. Note that as certain
symmetries of the infinite system are broken in the finite nucleus,
the above set of integral equations becomes a complicated one.  For
example, the lack of translational invariance induces the mixing
between various Lorentz structures ({\it e.g.,} timelike and
spacelike). Further, as the mean-field ground state is not isospin
symmetric, isoscalar and isovector modes will also get mixed.

\section{Results}
\label{sec:results}

\begin{figure}[ht]
\vspace{0.50in}
\includegraphics[width=5in,angle=0]{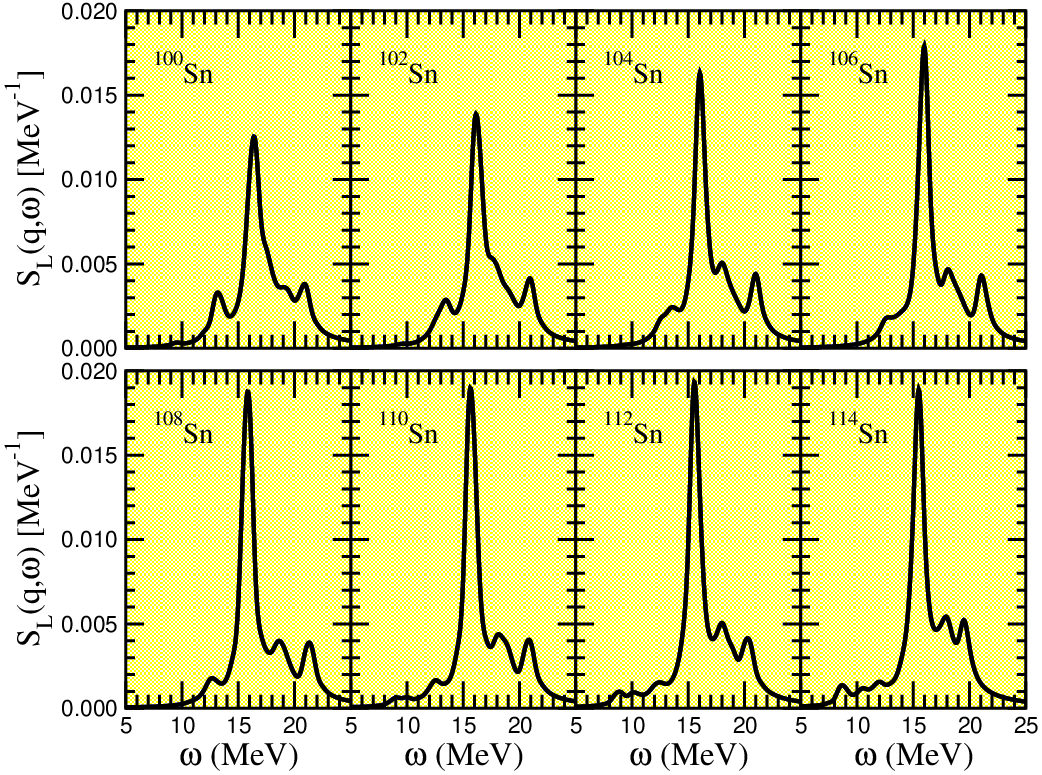}
\caption{Distribution of isovector dipole strength for all 
         neutron-even tin isotopes from ${}^{100}$Sn to
         ${}^{114}$Sn at a momentum transfer of 
	 $q\!=\!0.018~{\rm fm}^{-1}$. All calculations include
         a small artificial width of $\eta\!=\!0.5$~MeV and use
         the FSUGold parameter set~\cite{Todd-Rutel:2005fa}.}
\label{Fig2}
\end{figure}

\begin{figure}[ht]
\vspace{0.50in}
\includegraphics[width=5in,angle=0]{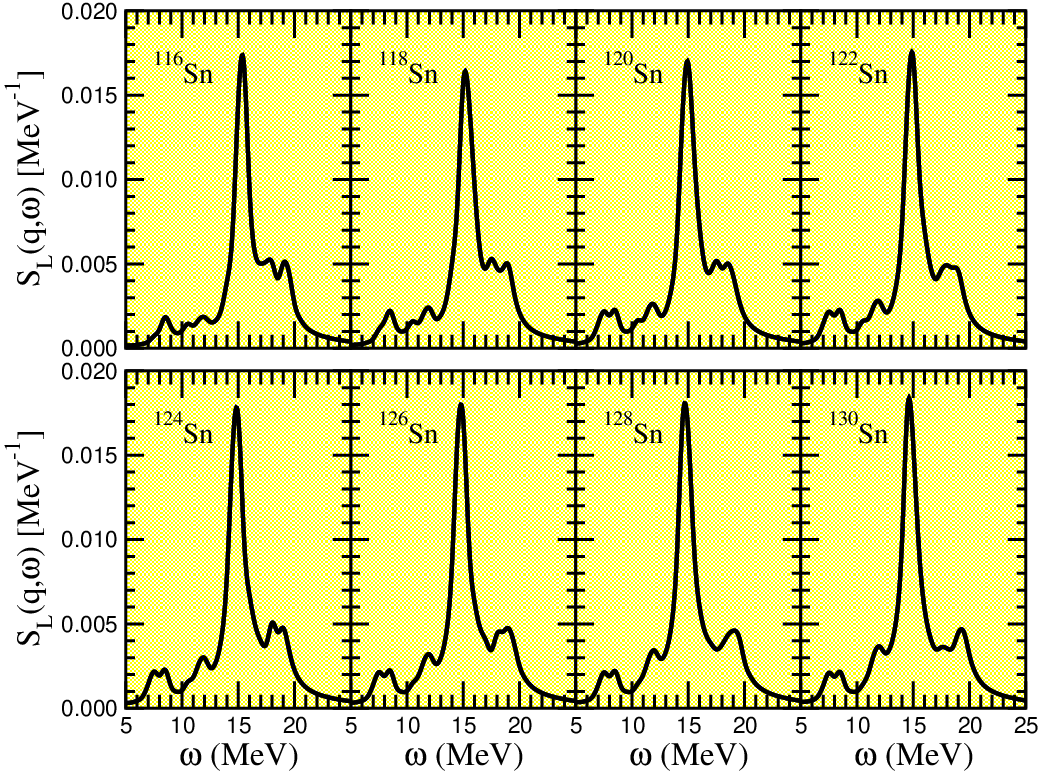}
\caption{Distribution of isovector dipole strength for all 
         neutron-even tin isotopes from ${}^{116}$Sn to
         ${}^{130}$Sn at a momentum transfer of 
	 $q\!=\!0.018~{\rm fm}^{-1}$. All calculations include
         a small artificial width of $\eta\!=\!0.5$~MeV and use
         the FSUGold parameter set~\cite{Todd-Rutel:2005fa}.}
\label{Fig3}
\end{figure}

We start this section by displaying in Figs.~\ref{Fig2}-\ref{Fig3} the
distribution of isovector dipole strength for various members of the
Sn-isotopic chain. (The distribution of strength in ${}^{132}$Sn is
shown in Fig.~\ref{Fig6}). In all cases the nuclear response is
reported at the small momentum transfer of $q\!=\!0.018~{\rm fm}^{-1}$
and includes a small artificial width of $\eta\!=\!0.5$~MeV. An
artificial width is included to resolve individual bound-state
transitions; due to the non-spectral character of our calculation,
particle-escape widths are computed exactly within the model.

%%%%%%%%%%%%%%%%%%%%%%%%%%%%%%%%%%%%%%%%%%%%%%%%%%%%%%%%%%%%%%%%%%%%%%
%%%%%%%%%%%%%%%%%%%%%%%%%%%%%%%%%%%%%%%%%%%%%%%%%%%%%%%%%%%%%%%%%%%%%%
%%%%%%%%%%%%%%%%%%%%%%%%%%%%%%%%%%%%%%%%%%%%%%%%%%%%%%%%%%%%%%%%%%%%%%

The large collective structure in the $\omega\!\sim\!15\!-\!16$~MeV
region corresponds to the quintessential nuclear mode: the isovector
giant dipole resonance (GDR). For medium-to-heavy nuclei, this
collective vibration represents a coherent oscillation of all protons
against all neutrons and is well-developed along the whole isotopic
chain~\cite{Bertsch:1994,Harakeh:2001}.  As is characteristic of these
collective excitations, a large fraction of the energy-weighted sum
rule is exhausted by this one resonance. But not all! The development
of low-energy ($\omega\!\sim\!7\!-\!9$~MeV) dipole strength with
increasing neutron number is clearly discerned in
Figs.~\ref{Fig2}-\ref{Fig3}.  The progressive addition of ``valence''
neutrons---those occupying the $1g^{7/2}$, $2d^{5/2}$, $2d^{3/2}$,
$3s^{1/2}$, and $1h^{11/2}$ orbitals (in that precise order)---results
in a well developed, albeit small, low-energy resonance. This
oscillation of the excess neutrons---the neutron skin---against the
isospin-symmetric core has been dubbed the ``pygmy dipole resonance''
(PDR). In addition to the full distribution of isovector dipole
strength, some of its moments have been tabulated in
Table~\ref{Table3}. To do so, an ad-hoc division was made at
10~MeV---with the PDR comprising the $5\!-\!10$~MeV low-energy region
and the GDR the $10\!-\!25$~MeV high-energy region. Note that only PDR
moments that account for at least one percent of the energy-weighted
sum rule are tabulated.

%%%%%%%%%%%%%%%%%%%%%%%%%%%%%%%%%%%%%%%%%%%%%%%%%%%%%%%%%%%%%%%%%
\begin{table}[ht]
\begin{tabular}{|c|c|c|c|c|c|c|c|c|}
 \hline
   & \multicolumn{3}{ c|}{Pygmy Dipole Resonance (PDR)} &
     \multicolumn{3}{ c|}{Giant Dipole Resonance (GDR)} &
     \multicolumn{2}{ c|}{PDR/GDR} \\
\hline
   A & $m_{0}~(10^{-2})$ & $m_{1}~(10^{-2}~{\rm MeV})$ & $E_{c}$~(MeV) 
     & $m_{0}~(10^{-2})$ & $m_{1}~({\rm MeV})$         & $E_{c}$~(MeV) 
     & $m_{0}~(\%)$ & $m_{1}~(\%)$ \\
\hline
100 & ---  & ---  & ---  & 4.58 & 0.79 & 17.22 &  ---  & ---  \\
102 & ---  & ---  & ---  & 4.70 & 0.81 & 17.17 &  ---  & ---  \\
104 & ---  & ---  & ---  & 4.81 & 0.82 & 17.11 &  ---  & ---  \\
106 & ---  & ---  & ---  & 4.91 & 0.84 & 17.06 &  ---  & ---  \\
108 & ---  & ---  & ---  & 5.01 & 0.85 & 17.00 &  ---  & ---  \\
110 & 0.14 & 1.23 & 8.51 & 5.12 & 0.86 & 16.85 &  2.81 & 1.42 \\
112 & 0.21 & 1.81 & 8.51 & 5.25 & 0.88 & 16.69 &  4.05 & 2.06 \\
114 & 0.27 & 2.30 & 8.49 & 5.37 & 0.89 & 16.52 &  5.08 & 2.60 \\
116 & 0.38 & 3.12 & 8.30 & 5.48 & 0.89 & 16.31 &  6.87 & 3.49 \\
118 & 0.47 & 3.86 & 8.21 & 5.60 & 0.90 & 16.10 &  8.40 & 4.28 \\
120 & 0.62 & 4.92 & 7.94 & 5.70 & 0.91 & 15.90 & 10.86 & 5.42 \\
122 & 0.62 & 4.89 & 7.93 & 5.78 & 0.92 & 15.88 & 10.66 & 5.33 \\
124 & 0.62 & 4.90 & 7.93 & 5.87 & 0.93 & 15.86 & 10.53 & 5.27 \\
126 & 0.61 & 4.84 & 7.92 & 5.95 & 0.94 & 15.84 & 10.27 & 5.14 \\
128 & 0.61 & 4.81 & 7.91 & 6.03 & 0.95 & 15.80 & 10.08 & 5.05 \\
130 & 0.60 & 4.77 & 7.91 & 6.11 & 0.96 & 15.78 &  9.88 & 4.95 \\
132 & 0.60 & 4.77 & 7.90 & 6.20 & 0.98 & 15.74 &  9.73 & 4.89 \\
\hline
\end{tabular}
\caption{Various moments of the isovector dipole distribution.
         The division between low-energy (pygmy) and high-energy 
         (giant) dipole strength was made arbitrarily at 10~MeV.}
\label{Table3}
\end{table}
%%%%%%%%%%%%%%%%%%%%%%%%%%%%%%%%%%%%%%%%%%%%%%%%%%%%%%%%%%%%%%%%%

\begin{figure}[ht]
\vspace{0.50in}
\includegraphics[width=5in,angle=0]{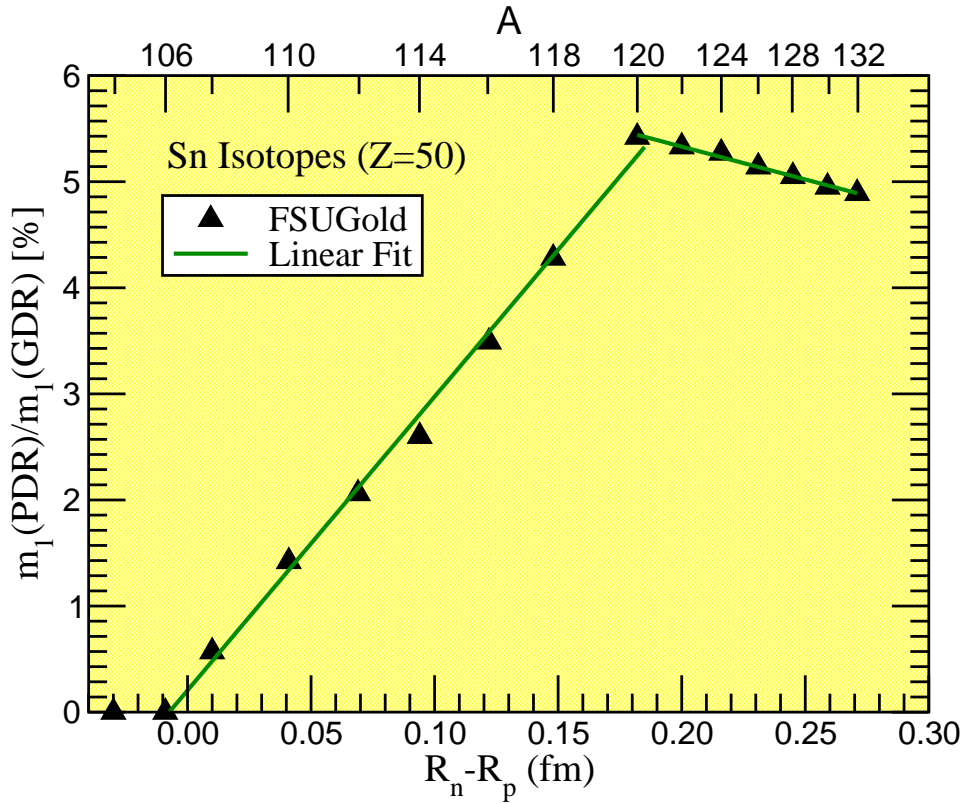}
\caption{Fraction of the energy-weighted sum rule contained in the
         low-energy region ($5\!-\!10$~MeV) relative to that in the 
         high-energy region ($10\!-\!25$~MeV) as a function of the 
         neutron skin of the various Sn-isotopes.}
\label{Fig4}
\end{figure}

We now turn to the first of the two questions posed in the
Introduction: is there a strong correlation between the development of
a neutron skin and the emergence of low-energy isovector dipole
strength in the tin isotopes? To answer this question we have plotted
in Fig.~\ref{Fig4} the fraction of the energy-weighted sum rule
contained in the PDR relative to that located in the GDR region
[${\cal M}_{1}\!\equiv\!m_{1}({\rm PDR})/m_{1}({\rm GDR})$] as a
function of the neutron skin ($R_{n}\!-\!R_{p}$). The same information
may be found in tabular form in Tables~\ref{Table1} and~\ref{Table3}.
Note that for those light isotopes for which the neutron skin is
negative, ${\cal M}_{1}$ is (as expected) vanishingly small.
Figure~\ref{Fig4} displays a strong (almost linear) correlation
between ${\cal M}_{1}$ and $R_{n}\!-\!R_{p}$ for $A\!\le\!120$.  This
lends support to the picture of the pygmy dipole resonance as an
oscillation of the excess neutrons in the skin against the
isospin-symmetric core. Yet as the neutron skin continues to increase
in going from ${}^{120}$Sn to ${}^{132}$Sn, a mild anti-correlation
actually develops. To elucidate these correlations it is useful to
call upon the single-particle (or mean-field) response.

%%%%%%%%%%%%%%%%%%%%%%%%%%%%%%%%%%%%%%%%%%%%%%%%%%%%%%%%%%%%%%%%%%%%%%
%%%%%%%%%%%%%%%%%%%%%%%%%%%%%%%%%%%%%%%%%%%%%%%%%%%%%%%%%%%%%%%%%%%%%%

\begin{figure}[ht]
\vspace{0.50in}
\includegraphics[width=6in,angle=0]{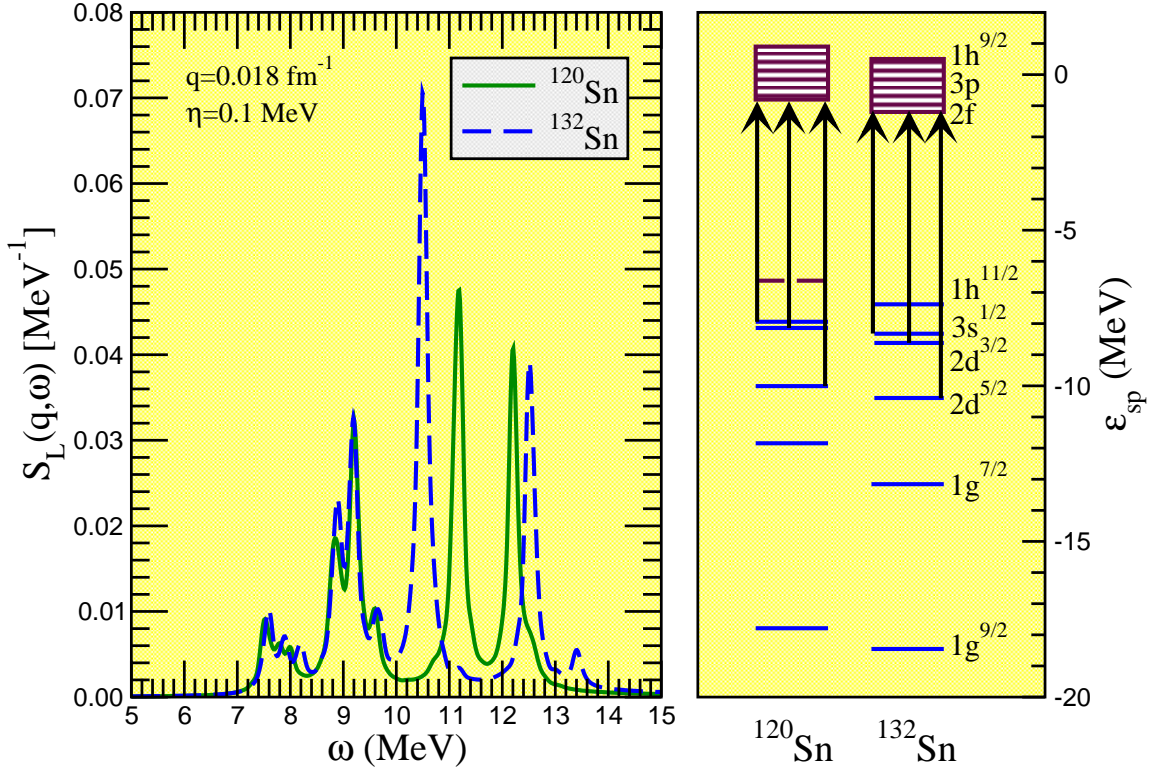}
\caption{The left-hand panel displays the mean-field distribution of 
         isovector dipole strength for ${}^{120}$Sn and ${}^{132}$Sn 
         at a momentum transfer of $q\!=\!0.018~{\rm fm}^{-1}$. The 
         calculations include a small artificial width of 
         $\eta\!=\!0.1$~MeV and use the FSUGold parameter 
         set~\cite{Todd-Rutel:2005fa}. The right-hand panel depicts 
         the allowed neutron transitions in the low-energy range.}
\label{Fig5}
\end{figure}

\begin{figure}[ht]
\vspace{0.50in}
\includegraphics[width=5in,angle=0]{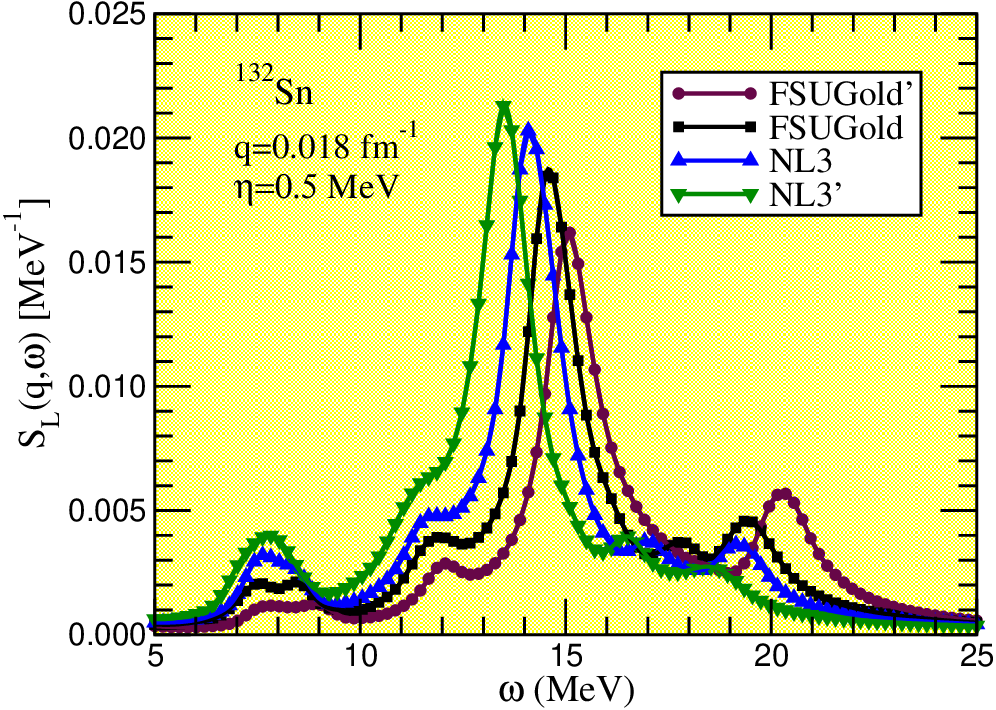}
\caption{Model dependence of the distribution of isovector dipole 
         strength in ${}^{132}$Sn at a momentum transfer of 
	 $q\!=\!0.018~{\rm fm}^{-1}$ and with a small artificial 
	 width of $\eta\!=\!0.5$~MeV.} 
\label{Fig6}
\end{figure}

The mean-field (or single-particle) isovector dipole response of both
${}^{120}$Sn and ${}^{132}$Sn are displayed on the left-hand panel of
Fig.~\ref{Fig5}.  To resolve most individual particle-hole transitions
the artificial width has been reduced from $\eta\!=\!0.5$~MeV to
$\eta\!=\!0.1$~MeV. The arrows on the right-hand panel indicate all
relevant low-energy ($\omega\!\lesssim\!10$~MeV) neutron excitations;
no proton transitions are possible in this energy range. Some of these
excitations involve states that are not bound (such as the $2f^{5/2}$
and $1h^{9/2}$), yet they are close enough in the continuum to produce
relatively sharp peaks. Two sets of single-neutron transitions are
clearly discernible in the $\omega\!\sim\!7\!-\!8$~MeV and
$\omega\!\sim\!9\!-\!10$~MeV energy regions---with their mixing
resulting in the ``two-hump'' PDR structure observed in the RPA
response (see Fig.~\ref{Fig3}). What is significant, however, is that
while the $1h^{11/2}$ neutron orbital gets filled in going from
${}^{120}$Sn to ${}^{132}$Sn, the low-energy structure of the
single-particle response remains unchanged. This indicates that
high-angular momentum orbitals play a passive role in driving
low-energy transitions of low multipolarity. As the main GDR peak
becomes more collective with increasing neutron number (see
Table~\ref{Table3}), the fraction of the energy-weighted sum rule
contained in the PDR peak actually goes down as the neutron skin
continues to increase, thereby generating the weak anti-correlation
displayed in Fig.~\ref{Fig4}.

%%%%%%%%%%%%%%%%%%%%%%%%%%%%%%%%%%%%%%%%%%%%%%%%%%%%%%%%%%%%%%%%%%%%%%
%%%%%%%%%%%%%%%%%%%%%%%%%%%%%%%%%%%%%%%%%%%%%%%%%%%%%%%%%%%%%%%%%%%%%%

Having answered the first question posed in the Introduction, we now
turn to the second and final one, namely, could one use the recently
measured isovector dipole strength in ${}^{130}$Sn and
${}^{132}$Sn~\cite{Adrich:2005} to discriminate among theoretical
models that, while accurately calibrated, yield different predictions
for the neutron skin of heavy nuclei. To address this question we
introduce---in addition to FSUGold---three other parameter sets. One
of these is the NL3 parameter
set~\cite{Lalazissis:1996rd,Lalazissis:1999} used earlier to compute
ground-state properties of the Sn-isotopes (see
Table~\ref{Table1}). Although enormously successful, NL3 predicts
considerably larger neutron skins than FSUGold (see
Tables~\ref{Table1} and~\ref{Table4}). This is a result of its stiff
symmetry energy---a property that at present is poorly determined by
existent ground-state data.  The remaining two sets (labeled FSUGold'
and NL3'), although themselves not calibrated by ground-state data,
they are close ``descendants'' of FSUGold and NL3---aimed at further
softening the symmetry energy of the former and stiffening the
symmetry energy of the latter.  To generate these two extra sets we
followed a prescription first outlined in Ref.~\cite{Horowitz:2001ya}
that preserves the value of the symmetry-energy coefficient at a
density that is slightly below nuclear-matter saturation density. This
prescription ensures that the density dependence of the symmetry
energy is modified without compromising the success of the model in
reproducing well-constrained ground-state observables. By introducing
these parameter sets, we have been able to generate a wide range of
values for the neutron skin of ${}^{130}$Sn and ${}^{132}$Sn (see
Table~\ref{Table4}). Note, however, that in the case of FSUGold', a
minor rescaling (of less than 0.5\%) of the isoscalar-scalar mass
($m_{s}$) was required to ensure the same proton radius in
${}^{132}$Sn for all four models.

The distribution of isovector dipole strength for the exotic,
neutron-rich isotope ${}^{132}$Sn is displayed in Fig.~\ref{Fig6}.  As
in all previous figures, the nuclear response is displayed at the
small momentum transfer of $q\!=\!0.018~{\rm fm}^{-1}$ and includes an
artificial width of $\eta\!=\!0.5$~MeV. In addition, neutron skins,
centroid energies $E_{c}\!\equiv\!m_{1}/m_{0}$, and PDR-to-GDR $m_{1}$
ratios for both ${}^{130}$Sn and ${}^{132}$Sn are listed in
Table~\ref{Table4}.  Whenever possible, these observables are compared
against the recently available experimental
data~\cite{Adrich:2005}. Perhaps surprisingly, the centroid energy of
the PDR is largely insensitive to the neutron skin. Yet there is a
noticeable enhancement with increasing neutron skin of the fraction of
the energy-weighted sum rule located at low energies. Although the
experimental error bars are large, the data seems to disfavor models
with overly large neutron skins. We regard this result as an important
and gratifying consistency check.  In recent theoretical studies of
the IVGDR in ${}^{208}$Pb, it was already suggested that models with a
stiff symmetry energy---and thereby large neutron skins---underpredict
the location of the
peak~\cite{Vretenar:2003qm,Piekarewicz:2003br}. Not surprisingly, the
same trend is observed here: models with a stiff symmetry energy, such
as NL3 and even more so NL3', predict GDR centroid energies in both
${}^{130}$Sn and ${}^{132}$Sn that are low relative to experiment.

%%%%%%%%%%%%%%%%%%%%%%%%%%%%%%%%%%%%%%%%%%%%%%%%%%%%%%%%%%%%%%%%%
\begin{table}[ht]
\begin{tabular}{|c|c|c|c|c|c|}
\hline
   & & & PDR & GDR & PDR/GDR \\
\hline
  Nucleus & Model & $R_{n}\!-\!R_{p}$~(fm) & $E_{c}$~(MeV) 
          & $E_{c}$~(MeV) & $m_{1}~(\%)$ \\
\hline
${}^{130}$Sn 
& FSUGold' & 0.180 & 7.94 & 16.42 & 3.38 \\
& FSUGold  & 0.259 & 7.91 & 15.78 & 4.95 \\
& NL3      & 0.330 & 7.91 & 15.13 & 6.49 \\
& NL3'     & 0.415 & 7.84 & 14.47 & 8.70 \\
\hline
& Experiment & --- & 10.1(7) & 15.9(5) & 0.05(2) \\
\hline
\hline
 ${}^{132}$Sn 
& FSUGold' & 0.189 & 8.01 & 16.59 & 3.04 \\
& FSUGold  & 0.271 & 7.90 & 15.74 & 4.89 \\
& NL3      & 0.346 & 7.90 & 15.08 & 6.44 \\
& NL3'     & 0.432 & 7.85 & 14.36 & 8.69 \\
\hline
& Experiment & --- & 9.8(7) & 16.1(7) & 0.03(2) \\
\hline
\end{tabular}
\caption{Model dependence of the neutron skins, PDR centroid energies,
 GDR centroid energies, and PDR-to-GDR $m_{1}$ ratios for ${}^{130}$Sn 
 and ${}^{132}$Sn. Experimental data is from Ref.~\cite{Adrich:2005}.}
\label{Table4}
\end{table}
%%%%%%%%%%%%%%%%%%%%%%%%%%%%%%%%%%%%%%%%%%%%%%%%%%%%%%%%%%%%%%%%%

We conclude this section with a comment addressing the apparent
discrepancy between theory and experiment on the location of the PDR
centroid energy. We find a substantial amount of low-energy strength
in bound excitations, an identification that is in agreement with
earlier studies~\cite{Paar:2002gz,Sarchi:2004}. For example, all
mean-field strength located in the $\omega\!\lesssim\!8$~MeV region
corresponds to bound-state transitions of the form
$3s^{1/2}\!\rightarrow\![3p^{3/2}, 3p^{1/2}]$;
$2d^{3/2}\!\rightarrow\![3p^{3/2}, 3p^{1/2}]$, and so on (see
Fig.~\ref{Fig5}). Further, most (although not all) of the strength
found in the $\omega\!\simeq\!8.5\!-\!10$~MeV region also involves
transitions to bound states. Yet experimentally, only isovector dipole
strength above the one-neutron separation energy was
detected~\cite{Adrich:2005}. Naturally, this tends to shift the
centroid energy of the PDR to higher energies. In an effort to
simulate the experimental conditions, we have turned off the
artificial width in our calculations (from
$\eta\!=\!0.5\!\rightarrow\!0$~MeV), thereby eliminating all
bound-state transitions. Implementing this procedure results in a
shift of the PDR centroid energy in ${}^{132}$Sn from
$E_{c}\!=\!7.90$~MeV to $E_{c}\!\simeq\!9.40$~MeV, well within
experimental error.

%%%%%%%%%%%%%%%%%%%%%%%%%%%%%%%%%%%%%%%%%%%%%%%%%%%%%%%%%%%%%%%%%%%%%%
%%%%%%%%%%%%%%%%%%%%%%%%%%%%%%%%%%%%%%%%%%%%%%%%%%%%%%%%%%%%%%%%%%%%%%

\section{Conclusions}
\label{sec:conclusions}

Motivated by the first experimental measurement of isovector dipole
strength in the two exotic, neutron-rich isotopes ${}^{130}$Sn and
${}^{132}$Sn, the theoretical response along the full (neutron-even)
isotopic chain was studied via a relativistic MF+RPA approach.
Particular emphasis was placed on the pygmy dipole resonance, an
oscillation of the excess neutrons in the skin against the
isospin-symmetric core. Our study was shaped by two underlying
questions: (a) is there a correlation between the development of a
neutron skin and the emergence of low-energy isovector dipole
strength? and (b) can the experimental data be used to discriminate
among models that, while accurately calibrated, predict different
values for the neutron skin of heavy nuclei?

We found that the answer to the first question was not unique. While a
strong linear correlation between the neutron skin and the fraction of
the energy-weighted sum rule at low energies was indeed observed, a
mild anti-correlation actually developed beyond ${}^{120}$Sn.  The
emergence of this anti-correlation was attributed to the $1h^{11/2}$
neutron orbital. While this orbital contributes significantly to the
size of the neutron skin, its large angular momentum hinders its
participation in low-energy transitions of low multipolarity.

The answer to the second question yielded various insights. First, the
centroid energy of the PDR was found to be insensitive to the density
dependence of the symmetry energy. That is, theoretical models that
predict neutron skins in ${}^{132}$Sn that vary by more than a factor
of two, yield centroid energies within 2\% of each other. Yet these
narrowly-spread centroid energies are about 2 MeV lower than
experiment. This discrepancy was attributed to the fact that the
observed experimental strength lies above the one-neutron separation
energy. To simulate this experimental condition, we suppressed all
bound-state transitions by turning off the artificial width from our
calculations. In accordance with experimental observations, the
centroid energy increased by 1.5~MeV. Finally, the fraction of the
energy-weighed sum rule exhausted by the PDR was observed to increase
sharply with increasing neutron skin, from 3\% for the model with the
softest symmetry energy to almost 9\% for the model with the stiffest
one. In spite of the relatively large experimental error bars, the
data seems to disfavor models with an overly stiff symmetry
energy. This result adds to the already large body of evidence
supporting a symmetry energy with a soft density
dependence~\cite{Todd-Rutel:2005fa,Li:2005sr,Chen:2005ti,Shetty:2005qp}

We conclude with a few comments on the impact of the pygmy dipole
resonance on various astrophysical phenomena. As already mentioned,
the systematics of this mode may be used to constrain the density
dependence of the symmetry energy, a property that has a strong impact
on a variety of neutron-star properties, such as its composition,
radius, and cooling
mechanism~\cite{Horowitz:2000xj,Horowitz:2001ya,Horowitz:2002mb}.
Further, the presence of low-energy dipole strength in neutron-rich
nuclei significantly enhances the cross section for the radiative
capture of low-energy ($\sim 10$~MeV)
neutrons~\cite{Goriely:2002cx,Goriely:2003rz}, a process of
fundamental importance to the creation of the heavy elements by means
of the rapid neutron capture process~\cite{Goriely:1998}.  Finally, the
existence of a pygmy dipole resonance in neutron-rich nuclei may aid
the supernovae explosion mechanism. In a supernovae explosion 99\% of
the energy of the collapse is radiated away in neutrinos.  Neutrinos
interact strongly with neutrons because of the large weak vector
charge of the neutron. Supernovae neutrinos may then couple strongly
to the neutron-rich skin and excite the low-energy modes of the many
exotic nuclei (``pasta'') present in this explosive
environment~\cite{Horowitz:2005zb}.  This may allow for a significant
energy transfer to the nuclear medium, potentially reviving the
stalled supernovae shock.

\smallskip
This work was supported in part by DOE grant DE-FG05-92ER40750.

\vfill\eject
\bibliography{ReferencesJP}

\end{document}